\documentclass[smallabstract,smallcaptions]{dccpaper}

\usepackage{epsfig}
\usepackage{citesort}
\usepackage{amsmath}
\usepackage{amssymb}
\usepackage{color}
\usepackage{url}

\newlength{\figurewidth}
\newlength{\smallfigurewidth}

\begin{document}

\title
{\large
\textbf{An experimental sorting method\\ for improving metagenomic data encoding}
}

\author{%
Diogo Pratas$^{1,2,3}$ and Armando J. Pinho$^{1,2}$\\[0.5em]
{\small\begin{minipage}{\linewidth}\begin{center}
\begin{tabular}{ccc}
$^1$IEETA/LASI - Institute of Electronics and Informatics Engineering of Aveiro\\
$^2$DETI - Department of Electronics, Telecommunications and Informatics\\
University of Aveiro, 3810-193 Aveiro, Portugal, \url{{pratas,ap}@ua.pt} \\
$^3$DoV - Department of Virology, University of Helsinki, 00014 Helsinki, Finland\\
~
\end{tabular}
\end{center}\end{minipage}}
}

\maketitle
\thispagestyle{empty}

\begin{abstract}
Minimizing data storage poses a significant challenge in large-scale metagenomic projects. In this paper, we present a new method for improving the encoding of FASTQ files generated by metagenomic sequencing. This method incorporates metagenomic classification followed by a recursive filter for clustering reads by DNA sequence similarity to improve the overall reference-free compression. In the results, we show an overall improvement in the compression of several datasets. As hypothesized, we show a progressive compression gain for higher coverage depth and number of identified species. Additionally, we provide an implementation that is freely available at \url{https://github.com/cobilab/mizar} and can be customized to work with other FASTQ compression tools.
\end{abstract}

\Section{Introduction}

Metagenomics is the study of genetic material collected from environmental samples, providing insights into the diversity, composition, and functional potential of microbial communities in various ecosystems. This includes permeates extreme ecosystems, such as uranium mines \cite{covas2017pedobacter}, intricate landscapes of both soft and hard tissues \cite{pyoria2023unmasking,toppinen2020landscape}, vital food crops \cite{qi2022haplotype}, expansive marine habitats \cite{katsanevakis2016mapping}, or in enigmatic depths of subterranean ecosystems \cite{cowan2015metagenomics}.

Metagenomics data is the digital material recovered from these samples by a method called sequencing, specifically using a format called FASTQ. The FASTQ format contains multiple reads, usually millions or billions. FASTQ reads are constituted by three channels of information: headers, DNA sequences, and quality-scores. In each read, DNA sequences and quality-scores must have the same number of characters, usually 150 or 250 characters. More recently, the technology of long reads is generating reads with several thousand or tens of thousand symbols, but at the expense of slightly less accuracy. The DNA sequence expresses the interpretation of the sequencing machine of a piece of the genetic material, while the quality-scores represent the quality in which the sequence has been sequenced (analogous to a confidence value). The headers (sometimes called identifiers) are specially used to know if the reads are paired (a correlation between pairs of reads that can extend in some way the size of the DNA sequences) or single reads and contain an identification to preserve the uniqueness of the reads. 

To speedup the sequencing process, the DNA is sequenced in parallel (cloning and reading) which generates a FASTQ file with reads in somewhat random order. These small DNA sequences can overlap given the random sequencing process and may have non sequenced parts. Fortunately, to reduce these holes and to increase the certainty in the sequencing process, the same regions are usually sequenced several times, but not necessarily starting at the same position. This property creates reads with the same DNA sub-sequences, where the average total number of a given DNA symbol that has been sequenced at a certain position of the genome depends on the coverage depth. Usually, the average coverage depth varies between 5 and 300, where a coverage depth of 30 (for human sequencing) is considered of high-quality. For smaller genome species, higher depth coverages are often, specially if enrichment processes have been applied in the sequencing process.

For some applications using paired-end reads, such as genome reconstruction, the information of the link between each forward and reverse read is required, while the order in which each pair appears is not mandatory. However, since the header contains the same string, with the exception at the end of a digit to describe if the read is in forward or reverse, grouping these pairs is trivial and does not require much information than a simple prefix-match program. Therefore, for most applications, the order of the reads can be neglected without compromising its purpose.

The compression of these data has widely been addressed using general-purpose data compressors. These compressors are prepared to deal with all kinds of finite data and some of them can be very fast and already be included, by default, in the operating systems. However, for compressing specific data, they can be much less compression efficient, especially in the case of FASTQ data. On the other hand, specialized FASTQ data compressors can take into account the different information channels, as well as to use methods more appropriate for each of the information channels. Additionally, they can model specific characteristics of DNA sequences, such as inverted repeats \cite{hosseini2016survey}.

There are many lossless FASTQ data compressors, for example, SCALCE \cite{hach2012scalce}, quip \cite{jones2012compression}, Fqzcomp \cite{bonfield2013compression}, DSRC and DSRC2 \cite{roguski2014dsrc}, ORCOM \cite{grabowski2015disk}, LEON \cite{benoit2015reference},  LW-FQZip2 \cite{huang2017lw}, HARC \cite{chandak2018compression}, FaStore \cite{roguski2018fastore}, SPRING \cite{chandak2019spring}, FastqCLS \cite{lee2022fastqcls}, among others. From these, only a few purely explore clustering or sorting the order of the reads for improving the compression \cite{wan2010sorting}, for example, HARC \cite{chandak2018compression}, FaStore \cite{roguski2018fastore}, and FastqCLS \cite{lee2022fastqcls}. However, as far as we know, none of the methods incorporate metagenomic classification and target metagenomic samples.

In this article, we provide a compression-based method that explores metagenomic classification and recursive filtering by similarity for sorting the order of the reads and increase the overall compression. Our main objective is to decrease the size of FASTQ files from metagenomics, without loss of information, while neglecting only the information associated to the order of which the reads appear.

Our main hypothesis stands for the successful ordering of FASTQ reads, through the approximation of similar classified DNA sequences using compression-based methods, in order to increase the overall FASTQ compression. Notice that the FASTQ headers and quality-scores can negatively affect the compression capacity when changing the order of the reads and, therefore, the overall compression gain must be superior to this eventual loss. In our second hypothesis, we argue that the compression gain using the proposed methodology is superior even if we store the information to transform the data to its original order (completely lossless). Our third hypothesis stands for the progressive increase in the compression gain for higher coverage depth and number of identified species, when our method is used.

In the next section, we present the method along with its implementation. Then, we present the results for validating our hypothesis and, finally, draw some conclusions. 

\Section{Method}

We propose a method that uses metagenomic classification for finding the sequences that are more probable to be contained in the sample, followed by recursive read filtering for clustering similar reads according to the select references, and compression using the sorted file, for improving the compression of FASTQ metagenomic data. The proposed method is designed for one-time compression followed by multiple end-users' decompression. It only necessitates complete computation of the sorting process during compression, enabling swift and resource-efficient decompression for all end-users.

\begin{figure*}[!h]
\centerline{\includegraphics[width=15.2cm]{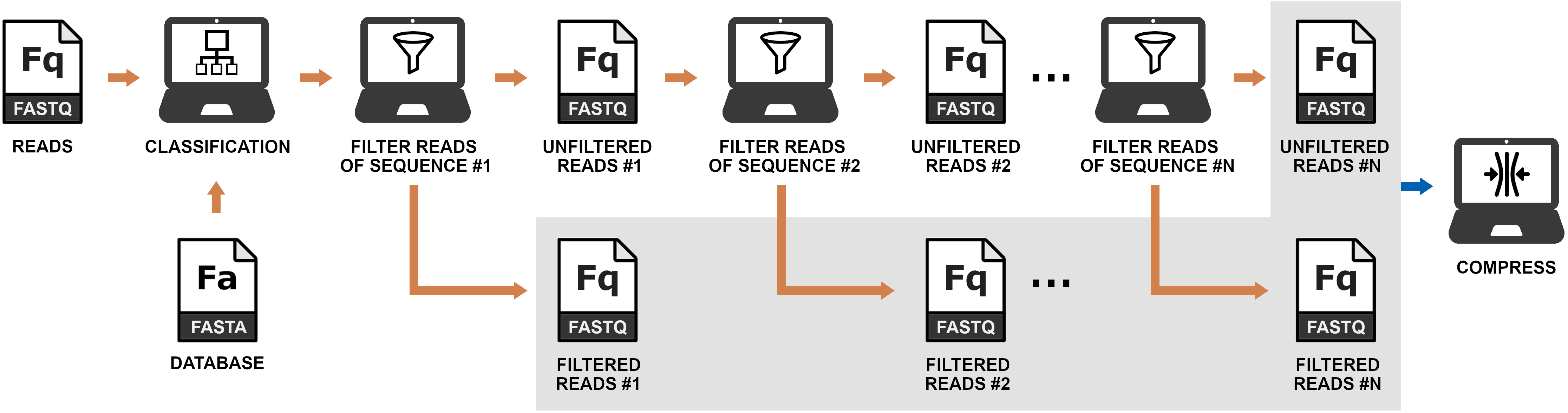}}
\caption{Architecture of the proposed methodology. }\label{method}
\end{figure*}

Figure~\ref{method} depicts the architecture of the proposed method, where each component is explained in the following subsections, namely the process of metagenomic classification, reads filtering, and reads compression.

\subsection*{Metagenomic classification}

In metagenomics, the DNA from several organisms is sequenced and presented conjointly. Knowing which organisms constitute a sample requires a database and a classification task. 

The creation of a database requires to download reference genome sequences, for example, from public repositories, with the disadvantage of requiring substantial disk space for storing the database, especially if metagenomic sequencing contains unknown species, where its composition can only be marginally guessed given frequency patterns from many classification of sets of FASTQ reads.

The classification task, which requires the previous creation of the database, is a key process in the proposed methodology. The current reference-free FASTQ data compressors do not take classification into account, employing only models that try to cope and adapt to the nature of unknown sequences in the metagenomic sample. 

For increasing the capability of bringing closer similar DNA sequences of FASTQ reads, that in principle share higher similarity according to higher coverage depth, we employ compression-based schemes \cite{pratas2018metagenomic,pratas2018metagenomic2}. 

Consider a set of reference sequences $X$ constituted by $x_1, x_2, ..., x_n$ and a set of FASTQ read sequences $Y$ constituted by $y_1, y_2, ..., y_m$. For convenience, we fix the alphabet $\Sigma$ of both $X$ and $Y$ to $\Sigma=\{A, C, G, T\}$. Therefore, symbols that fall outside this alphabet are considered a random symbol from $\Sigma$ (only in the analysis phase). 

The proposed method uses at its core data compression to estimate the similarity of the reads according to an existing reference DNA sequence, $x_i$, where reference sequences that have higher similarity are more probable of being present on the sequences sample. Specifically, the method loads into several context and substitution tolerant context models of several depths the DNA sequences present on FASTQ reads of a given file. Then, freezes these models and, for each DNA reference sequence, estimates the number of bits required to compress it, $C(x_i||Y)$. Notice that $C(x_i||Y)$ stands for the relative compression of $x_i$ given exclusively $y$ \cite{pinho2016authorship}. Then, the percentage of relative similarity, $S(x_i$), is calculated according to 
\begin{equation}
    S(x_i) = \left(1 - \frac{C(x_i||Y)}{|x_i| \log_2|\Sigma|}\right) \times 100,
\end{equation}
where $x_i$ is the reference sequence to be compressed, $Y$ the sequence with all the reads concatenated, $|x_i|$ the number of symbols in $x_i$, and $|\Sigma|$ the size of the alphabet of $x_i$.

Finally, the relative similarities, along with each associated reference sequence information, are sorted by ascending order. From this classification information, we take the most similar sequences above a certain threshold $\mathcal{T}_1$, namely $S(x_i) > \mathcal{T}_1$, and provide this information to the next phase, the reads filtering.

\subsection*{Reads filtering}

In this phase, the objective is to filter input FASTQ reads, $Y$, that have similarity higher than $\mathcal{T}_1$ for each reference sequence, $x_i$, followed by relative compression ($R$) lower than a given threshold, $\mathcal{T}_2$. For filtering these reads, analogous to the metagenomic classification phase, we apply data compression. Specifically, the proposed method loads into several context and substitution tolerant context models the highest similar reference sequence from $X$. Then, for each read, $y_j$ of the FASTQ metagenomic file, the number of bits needed to compress using exclusively the selected reference $x_i$ is calculated according to
\begin{equation}
    R(y_j) = \frac{C(y_j||x_i)}{|y_j|\log_2|\Sigma|}.
\end{equation}

According to Figure~\ref{method}, the reads where $R(y_j) \leq \mathcal{T}_2$ are redirected to a FASTQ file called filtered reads, otherwise to a FASTQ file called unfiltered reads. 

Then, the unfiltered reads are used as input to repeat the described process for the second highest similar reference sequence determined in the metagenomic classification phase. The process is again repeated until the number of reference sequences above $\mathcal{T}_1$ is reached or the number of unfiltered reads is zero.

The last sub-phase is constituted by the concatenation of all the filtered reads and the last set of unfiltered reads (if exists) is sent to a single file. In theory, reads with higher similarity should be more close, according to the identified reference sequence in the metagenomic classification phase. Notice that this final file contains the same number of characters than the original file.

\subsection*{Reads compression}

In principle, data compressors with larger memory model capacity will require larger FASTQ files for noticing compression gains. Additionally, this type of compressors require more computational resources, including time, which for a scenario of one compression on a server side and multiple downloads in client sides significantly reduces its usage. In a FASTQ file sorted by similarity, these large memory data compressors show less effectiveness because when similar blocks are closer the need for a larger memory model is reduced.

For this purpose, we choose three data compressors to incorporate the proposed method that are available by parameter selection, namely LZMA, Fqzcomp, JARVIS. LZMA \footnote{Available for free download at \url{https://tukaani.org/lzma}} is a general purpose data compressor. Fqzcomp \cite{bonfield2013compression} is a specific FASTQ data compressor that uses low computational resources and good compression capabilities. JARVIS (Version 3) \footnote{Development version is available for free download at \url{https://github.com/cobilab/jarvis3}} is an experimental genomic compressor based on JARVIS2 \cite{pratas2023jarvis2}. Both Fqzcomp and JARVIS use different models for compressing headers, DNA sequences, and quality-scores. 

\subsection*{Implementation}

The implementation of the proposed method (named MizaR) is done using the C/C++ language and Bash. This implementation is available for free download (GPv3 license) at \url{https://github.com/cobilab/MizaR} along with the scripts to generate all the results presented in this article.

\Section{Results}

In this section, we evaluate the performance of the proposed method using several datasets. These datasets are composed by real viral sequences in which the process of sequencing is simulated using different conditions. The reason to use sequencing simulation is to understand the dynamics that affect different sequencing coverage depth and number of input sequences.

For simulating the process of sequencing, we used the ART simulator \cite{huang2012art}. The ART simulator generates reads using the information of genome sequences that are feed using a multi-FASTA format. Notice that this simulator includes alterations to better approximate the sequencing process. The size of the reads was set to 150 base pairs, the mean size (-m) to 200 and the standard deviation (-s) to 10 while set for paired-end reads similar to a Illumina HiSeq 2500. The genome sequences were picked pseudo-randomly for each run from the viral database in use (also available at the repository). For each ART simulation, the depth sequencing coverage was set to a specific value and MizaR run with the default threshold parameters ($\mathcal{T}_1=50$ and $\mathcal{T}_2=0.5$). 

All the computations have been performed using a laptop computer running Linux Ubuntu 20.04 with Intel® Core™ i7-8650U CPU @ 1.90GHz × 8, 15.5 GiB RAM, and a disk of approximately 512 GB. 

\begin{figure*}[!h]
\centerline{\includegraphics[width=13cm]{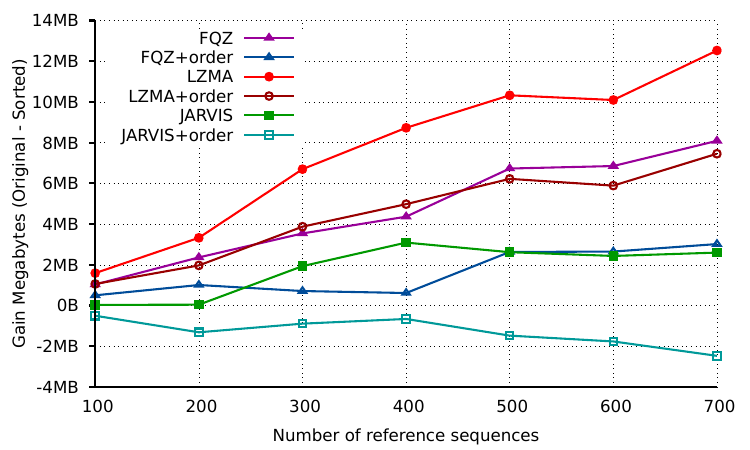}}
\caption{Compression gain in Megabytes (original compressed number of bytes minus the sorted compressed number of bytes) for the Fqzcomp, JARVIS, and LZMA compressors while varying the number of reference sequences used in the sequencing simulation.  }\label{sequences}
\end{figure*}

Figure~\ref{sequences} depicts the compression gain for different compressors (Fqzcomp, JARVIS, LZMA) with and without preserving the order of the reads using the simulation of sequencing from real viral reference sequences while using a coverage depth of 50 and varying the number of reference sequences from 100 to 700.

As it can be seen, there is a progressive increase in the compression gain as the number of reference sequences increases. By compression gain, we mean the Megabytes of the compressed original file minus the compressed sorted file using the same data compressor. There is one point below the average increase (for 600 reference sequences) because the pseudo-random selection was performed from nearly 3k sequences and the higher values have higher probability of contain similar reference sequences which harm the compression gain.

Figure~\ref{sequences} also depicts that the general-purpose compressor LZMA obtains higher compression gains compared to specialized FASTQ compressors (Fqzcomp and JARVIS). Moreover, LZMA compression gain was similar to the specialized compressor Fqzcomp if we would need to account for the information required to represent the order of the reads and using the worst model, approximated by the Stirling's formula, $y^n\log_2(y^n) - y^n \log_2e$, where $y^n$ stands for the number of reads of $y$. 

Globally, these gains represent, for example, in 700 reference sequences, approximately 3.7\% of improvement using the sorting methodology (computational time of sorting with 8 threads was 93 minutes) compared to not using the sorting methodology; assuming the percentage gain as $(1-C_s/C_u)\times 100$, where $C_s$ and $C_u$ stand for the compression of the sorted and unsorted files, respectively. For the same example, considering the worst case for preserving the order, the compression percentage gain is approximately 2.2\%. 

The JARVIS compressor seems to achieve a lower gain than Fqzcomp and LZMA. This is justified by the memory model that is larger than in the other compressors. For starting to notice higher gains in JARVIS, the size of the file must be larger. Usually, larger files are the most frequent case, where files have one order of magnitude large than those used here. However, because there were many computations to perform, we used smaller examples.

\begin{figure*}[!h]
\centerline{\includegraphics[width=13cm]{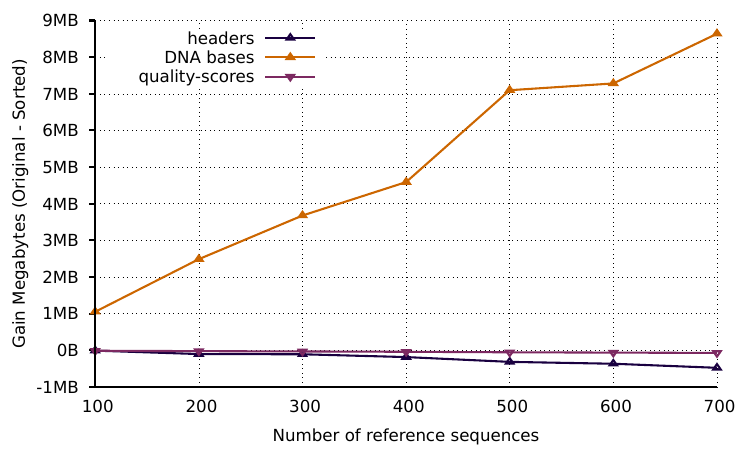}}
\caption{Compression gain in Megabytes (original compressed number of bytes minus the sorted compressed number of bytes) for the Fqzcomp compressor while varying the number of reference sequences used in the sequencing simulation.  }\label{channels}
\end{figure*}

Figure~\ref{channels} depicts the gains for each of the information channels (headers, DNA sequences, and quality-scores) while using Fqzcomp. We can notice a higher gain in the DNA sequence channel (as expected) and a small negative gain in the headers, while the quality-scores remain approximately the same. 

\begin{figure*}[!h]
\centerline{\includegraphics[width=13cm]{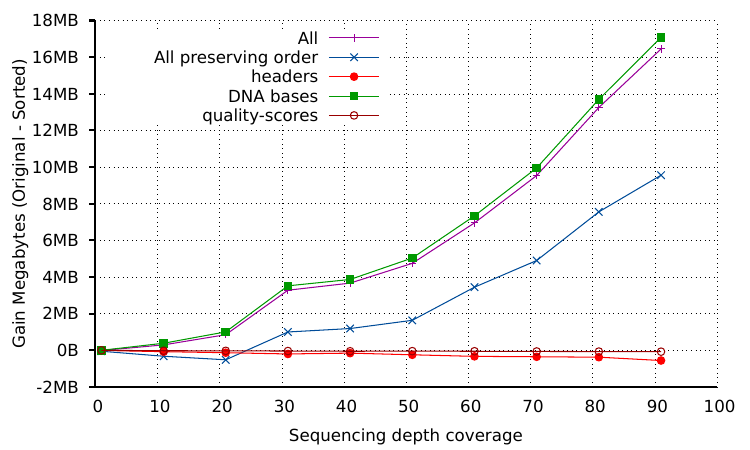}}
\caption{Compression gain in Megabytes (original compressed number of bytes minus the sorted compressed number of bytes) for the Fqzcomp compressor while varying the sequencing coverage depth used in the sequencing simulation. }\label{coverage}
\end{figure*}

Accordingly, these values correspond to an average coverage depth of 50, which for enriched viral metagenomic samples is very low. For studying the impact of the coverage depth, we applied the same experience but now fixing the number of reference sequences to 500 and varying the coverage depth from 1 to 91. Figure~\ref{coverage} depicts this analysis, showing a gain increase for higher coverage depth.

Notice that MizaR depends on the diversity and quality of the database for metagenomic classification. The recursive filter will only group reads that share similarity above the defined threshold with the reference sequence in the database. Therefore, building a good reference sequence database is of extreme importance for increasing the compression gain capability of MizaR.

Additionally, MizaR is very flexible because it can be applied in other contexts, such as human genome sequencing. For this case, besides the usual bacterial and viral contamination in the database, the human genome can also be included in the reference sequences database. However, for improved compression, instead of having a reference for each chromosome, these chromosomal references can be separated into junks of, say, 5 MB. MizaR will assume different references sequences and will provide better resolution for such a large genome.

The final considerations are for the parameters, namely the $\mathcal{T}_1$ and $\mathcal{T}_2$ thresholds. Although MizaR already provides default values, $\mathcal{T}_1=50$ and $\mathcal{T}_2=0.5$, these may require adaptation according to the application. This flexibility can be an advantage, but it can also be a disadvantage because without expertise in the area it is difficult to know which parameters better work for each case. Nevertheless, an intuition can be given. For example, for $\mathcal{T}_1<50$ and a high number of similar references in the database, MizaR will filter some of these sequences and collect only a few reads, increasing the computational time without increasing much the compressing gain. For different values of $\mathcal{T}_2$, the reads, which share similarity with other reads corresponding to other genome species, will be grouped or missed. 

The choice of these parameters calls for a study which target different case scenarios. Moreover, the tests to compute Figure~\ref{sequences} used at most a file size less than 2 GB, which are only indicative of the true potential of MizaR, because higher gains are associated with larger files (for example $>20$ GB).

\Section{Conclusions}

In this paper, we presented MizaR, a compression-based method that uses metagenomic classification and recursive filtering for improving the compression of FASTQ metagenomic files, by changing the order of the reads through DNA sequence similarity. We have shown that our methodology is able to provide additional gains as the sequencing coverage depth increases and the number of reference sequences. We have shown that general-purpose compression methods, such as LZMA, are able to benefit more with this methodology.

\section*{Acknowledgements}

D.P. is funded by national funds through FCT – Fundação para a Ciência e a Tecnologia, I.P., under the Scientific Employment Stimulus - Institutional Call - reference CEECINST/00026/2018. We also acknowledge national funds through the FCT in the context of the project UIDB/00127/2020.

\Section{References}
\bibliographystyle{IEEEbib}

\begin{thebibliography}{10}

\bibitem{covas2017pedobacter}
Cl{\'a}udia Covas, T{\^a}nia Caetano, Andreia Cruz, Tiago Santos, Liliana Dias,
  Guenter Klein, Amir Abdulmawjood, Luis~M Rodr{\'\i}guez-Alcal{\'a},
  L{\'\i}gia~L Pimentel, Ana Gomes, et~al.,
\newblock ``{Pedobacter lusitanus sp. nov., isolated from sludge of a
  deactivated uranium mine},''
\newblock {\em International journal of systematic and evolutionary
  microbiology}, vol. 67, no. 5, pp. 1339--1348, 2017.

\bibitem{pyoria2023unmasking}
Lari Py{\"o}ri{\"a}, Diogo Pratas, Mari Toppinen, Klaus Hedman, Antti
  Sajantila, and Maria~F Perdomo,
\newblock ``{Unmasking the tissue-resident eukaryotic DNA virome in humans},''
\newblock {\em Nucleic Acids Research}, vol. 51, no. 7, pp. 3223--3239, 2023.

\bibitem{toppinen2020landscape}
Mari Toppinen, Diogo Pratas, Elina V{\"a}is{\"a}nen, Maria
  S{\"o}derlund-Venermo, Klaus Hedman, Maria~F Perdomo, and Antti Sajantila,
\newblock ``{The landscape of persistent human DNA viruses in femoral bone},''
\newblock {\em Forensic Science International: Genetics}, vol. 48, pp. 102353,
  2020.

\bibitem{qi2022haplotype}
Weihong Qi, Yi-Wen Lim, Andrea Patrignani, Pascal Schl{\"a}pfer, Anna
  Bratus-Neuenschwander, Simon Gr{\"u}ter, Christelle Chanez, Nathalie Rodde,
  Elisa Prat, Sonia Vautrin, et~al.,
\newblock ``{The haplotype-resolved chromosome pairs of a heterozygous diploid
  African cassava cultivar reveal novel pan-genome and allele-specific
  transcriptome features},''
\newblock {\em GigaScience}, vol. 11, pp. giac028, 2022.

\bibitem{katsanevakis2016mapping}
Stelios Katsanevakis, Fernando Tempera, and Heliana Teixeira,
\newblock ``{Mapping the impact of alien species on marine ecosystems: the
  Mediterranean Sea case study},''
\newblock {\em Diversity and Distributions}, vol. 22, no. 6, pp. 694--707,
  2016.

\bibitem{cowan2015metagenomics}
Don~A Cowan, Jean-Baptiste Ramond, Thulani~P Makhalanyane, and Pieter
  De~Maayer,
\newblock ``{Metagenomics of extreme environments},''
\newblock {\em Current opinion in microbiology}, vol. 25, pp. 97--102, 2015.

\bibitem{hosseini2016survey}
Morteza Hosseini, Diogo Pratas, and Armando~J Pinho,
\newblock ``{A survey on data compression methods for biological sequences},''
\newblock {\em Information}, vol. 7, no. 4, pp. 56, 2016.

\bibitem{hach2012scalce}
Faraz Hach, Ibrahim Numanagi{\'c}, Can Alkan, and S~Cenk Sahinalp,
\newblock ``{SCALCE: boosting sequence compression algorithms using locally
  consistent encoding},''
\newblock {\em Bioinformatics}, vol. 28, no. 23, pp. 3051--3057, 2012.

\bibitem{jones2012compression}
Daniel~C Jones, Walter~L Ruzzo, Xinxia Peng, and Michael~G Katze,
\newblock ``{Compression of next-generation sequencing reads aided by highly
  efficient de novo assembly},''
\newblock {\em Nucleic acids research}, vol. 40, no. 22, pp. e171--e171, 2012.

\bibitem{bonfield2013compression}
James~K Bonfield and Matthew~V Mahoney,
\newblock ``{Compression of FASTQ and SAM format sequencing data},''
\newblock {\em PloS one}, vol. 8, no. 3, pp. e59190, 2013.

\bibitem{roguski2014dsrc}
{\L}ukasz Roguski and Sebastian Deorowicz,
\newblock ``{DSRC 2—Industry-oriented compression of FASTQ files},''
\newblock {\em Bioinformatics}, vol. 30, no. 15, pp. 2213--2215, 2014.

\bibitem{grabowski2015disk}
Szymon Grabowski, Sebastian Deorowicz, and {\L}ukasz Roguski,
\newblock ``{Disk-based compression of data from genome sequencing},''
\newblock {\em Bioinformatics}, vol. 31, no. 9, pp. 1389--1395, 2015.

\bibitem{benoit2015reference}
Ga{\"e}tan Benoit, Claire Lemaitre, Dominique Lavenier, Erwan Drezen, Thibault
  Dayris, Raluca Uricaru, and Guillaume Rizk,
\newblock ``{Reference-free compression of high throughput sequencing data with
  a probabilistic de Bruijn graph},''
\newblock {\em BMC bioinformatics}, vol. 16, no. 1, pp. 1--14, 2015.

\bibitem{huang2017lw}
Zhi-An Huang, Zhenkun Wen, Qingjin Deng, Ying Chu, Yiwen Sun, and Zexuan Zhu,
\newblock ``{LW-FQZip 2: a parallelized reference-based compression of FASTQ
  files},''
\newblock {\em BMC bioinformatics}, vol. 18, pp. 1--8, 2017.

\bibitem{chandak2018compression}
Shubham Chandak, Kedar Tatwawadi, and Tsachy Weissman,
\newblock ``{Compression of genomic sequencing reads via hash-based reordering:
  algorithm and analysis},''
\newblock {\em Bioinformatics}, vol. 34, no. 4, pp. 558--567, 2018.

\bibitem{roguski2018fastore}
{\L}ukasz Roguski, Idoia Ochoa, Mikel Hernaez, and Sebastian Deorowicz,
\newblock ``{FaStore: a space-saving solution for raw sequencing data},''
\newblock {\em Bioinformatics}, vol. 34, no. 16, pp. 2748--2756, 2018.

\bibitem{chandak2019spring}
Shubham Chandak, Kedar Tatwawadi, Idoia Ochoa, Mikel Hernaez, and Tsachy
  Weissman,
\newblock ``{SPRING: a next-generation compressor for FASTQ data},''
\newblock {\em Bioinformatics}, vol. 35, no. 15, pp. 2674--2676, 2019.

\bibitem{lee2022fastqcls}
Dohyeon Lee and Giltae Song,
\newblock ``{FastqCLS: a FASTQ compressor for long-read sequencing via read
  reordering using a novel scoring model},''
\newblock {\em Bioinformatics}, vol. 38, no. 2, pp. 351--356, 2022.

\bibitem{wan2010sorting}
Raymond Wan and Kiyoshi Asai,
\newblock ``{Sorting next generation sequencing data improves compression
  effectiveness},''
\newblock in {\em 2010 IEEE International Conference on Bioinformatics and
  Biomedicine Workshops (BIBMW)}. IEEE, 2010, pp. 567--572.

\bibitem{pratas2018metagenomic}
Diogo Pratas and Armando~J Pinho,
\newblock ``{Metagenomic composition analysis of sedimentary ancient DNA from
  the Isle of Wight},''
\newblock in {\em 2018 26th european signal processing conference (EUSIPCO)}.
  IEEE, 2018, pp. 1177--1181.

\bibitem{pratas2018metagenomic2}
Diogo Pratas, Morteza Hosseini, Gon{\c{c}}alo Grilo, Armando~J Pinho, Raquel~M
  Silva, T{\^a}nia Caetano, Jo{\~a}o Carneiro, and Filipe Pereira,
\newblock ``{Metagenomic composition analysis of an ancient sequenced polar
  bear jawbone from Svalbard},''
\newblock {\em Genes}, vol. 9, no. 9, pp. 445, 2018.

\bibitem{pinho2016authorship}
Armando~J Pinho, Diogo Pratas, and Paulo~JSG Ferreira,
\newblock ``{Authorship attribution using relative compression},''
\newblock in {\em 2016 Data Compression Conference (DCC)}. IEEE, 2016, pp.
  329--338.

\bibitem{pratas2023jarvis2}
Diogo Pratas and Armando~J Pinho,
\newblock ``{JARVIS2: a data compressor for large genome sequences},''
\newblock in {\em 2023 Data Compression Conference (DCC)}. IEEE, 2023, pp.
  288--297.

\bibitem{huang2012art}
Weichun Huang, Leping Li, Jason~R Myers, and Gabor~T Marth,
\newblock ``{ART: a next-generation sequencing read simulator},''
\newblock {\em Bioinformatics}, vol. 28, no. 4, pp. 593--594, 2012.

\end{thebibliography}

\end{document}